\documentstyle[12pt,epsf,clbiba]{article}
\title{A Lattice-Gas with Long-Range Interactions Coupled to a Heat Bath}

\author{Jeffrey Yepez \\
{\small\it US Air Force, Phillips Laboratory, Hanscom Field, Massachusetts,
01731} }

\date{September 10, 1993}

\begin{document}

\maketitle

\baselineskip=18pt

\begin{abstract}

Introduced is a lattice-gas with long-range 2-body interactions.  An effective
inter-particle force
is mediated by momentum exchanges.  There exists the possibility of having both
attractive and
repulsive interactions using finite impact parameter collisions.  There also
exists an interesting
possibility of coupling these long-range interactions to a heat bath. A fixed
temperature heat
bath induces a permanent net attractive interparticle potential, but at the
expense of reversibility.
Thus the long-range dynamics is a kind of a Monte Carlo Kawasaki updating
scheme. The model
has a $P\rho T$ equation of state.  Presented are analytical and numerical
results for the a
lattice-gas fluid governed by a nonideal equation of state.  The model's
complexity is not much
beyond that of the FHP lattice-gas. It is suitable for massively parallel
processing and may be
used to study critical phenomena in large systems.
\end{abstract}

\pagenumbering{roman}

\pagenumbering{arabic}

\newcommand{\ea}{\mbox{${\bf \hat e_a}$}}
\newcommand{\e}{\mbox{${\bf \hat e}$}}
\newcommand{\r}{\mbox{${\bf r}$}}
\newcommand{\p}{\mbox{${\bf p}$}}
\newcommand{\x}{\mbox{${\bf x}$}}
\newcommand{\X}{\mbox{${\bf X}$}}
\newcommand{\vel}{\mbox{${\bf v}$}}

\newcommand{\ah}{\mbox{${\hat a}$}}
\newcommand{\ahd}{\mbox{${\hat a^\dagger}$}}
\newcommand{\n}{\mbox{${\hat n}$}}

\newcommand{\vac}{\mbox{$\mid 0 \rangle$}}
\newcommand{\one}{\mbox{${\bf 1}$}}
\newcommand{\mex}{\mbox{${\bf \hat{\chi}}$}}
\newcommand{\U}{\mbox{${\bf \hat{U}}$}}
\newcommand{\N}{\mbox{${\bf \hat{N}}$}}
\newcommand{\stream}{\mbox{${\bf \hat{S}}$}}
\newcommand{\ham}{\mbox{${\bf \hat{H}}$}}

\section{Introduction}

Nonideal fluids, with dynamics governed by {\it reversible} physical laws,
undergo phase
transitions.  This fact about fluids indicates the possibility that lattice-gas
fluid models, with
dynamics governed by {\it reversible} rules \cite{fredkin-82}, may also undergo
phase
transitions.  The Ising model is the most well known computational model with
an order-disorder
transition.  Reversible Ising models using energy bankers, in a microcanonical
ensemble, are
known \cite{toffoli-87,creutz-prl92}.  Yet it is an open question as to whether
or not there exists a
reversible lattice-gas model of a multiphase fluid.

In molecular dynamics one simulates a many-body system of particles with
continuous
interaction potentials where the particles have continuous positions and
momenta.  In lattice-gas
dynamics the particles' positions and momenta are discrete and motion is
constrained to a
spacetime lattice.  Interparticle potentials can be modeled by including
long-range interactions
in the lattice-gas dynamics with a discrete momentum exchange between
particles. The use of
momentum exchange was introduced by Kadanoff and Swift in a Master-equation
approach
\cite{kadanoff-jul67}.  The use of negative momentum exchanges in long-range
interactions was
first done in a lattice-gas model by Appert and Zaleski \cite{appert-90}.
Their nonthermal
model has a liquid-gas coexistence phase; there is a $P\rho$ equation of state.
   A method for
modeling interparticle potentials using only local interactions was introduced
by Chen {\it et al.}
\cite{chen-sep89}. They extended the FHP lattice-gas model\cite{frisch-86} by
adding a rest
particle state which has a certain Ising-like interaction with neighboring rest
particles; there is a
local configurational energy associated with the rest particles. Speed one
particles can change
into a rest state with a certain Boltzmann probability, $e^{-\beta \Delta E}$.
The inverse
transition is also possible. Chen {\it et al.} observed an order-disorder
transition; their system has
a nonideal equation of state.

The lattice-gas model with long-range interactions presented here can be viewed
as a finite
temperature extension of Appert and Zaleski's zero temperature model.  The
first ingredient
added is repulsive long-range collisions. Both discrete negative and positive
momentum
exchanges occur between particles.   The second ingredient added is a finite
temperature
heat-bath.  It is possible to bias the finite impact parameter collisions so
there {\it is} a net
attractive interparticle potential.  This is done by coupling the long-range
collisions to a heat
bath --- attractive collisions cause a transition from a high potential energy
state to a low one
and emit units of heat whereas repulsive collisions cause the opposite
transition and absorb
heat.  When a disordered lattice-gas state is in contact with a low temperature
bath, liquid and
gas phases emerge.  The probability of a momentum exchange event depends on the
heat bath
temperature.

When the lattice-gas is in contact with a fixed temperature heat-bath, the
dynamics is
irreversible. The phase separation endures permanently in time. This
lattice-gas automaton now
has finite temperature equilibria. Its nonideal $P\rho T$ equation of state is
derived and
compared to numerical simulation.  Boltzmann probabilities are ``input'' into
the model by setting
the heat-bath density, and in essence is similar to  the local method of Chen
{\it et al.}
\cite{chen-sep89}.   Like the Ising model it is very simple and has practical
theoretical and
computational value.  Its advantage over the Ising model is its momentum
conservation and can
therefore be used to view the kinetics, even near the critical point.

When the lattice-gas dynamics is strictly reversible, there exists an inherent
limitation that the
phase separation process can occur only for a short period of time.   The
lattice-gas fluid quickly
becomes a neutral fluid, with finite impact parameter collisions. So added to
the usual FHP type
on-site collisions are an equivalent set of finite impact parameter collisions.
 Balancing the
interactions ensures detailed balance, and in the context of the multiphase
model presented
below, this is like an infinite temperature limit.   The appendix contains a
description of the
reversible lattice-gas with balanced attractive and repulsive collisions and a
numerical result
illustrating the characteristic transient time.

This paper is organized into three main sections.
\S\ref{lattice-gas-automata} very briefly
describes the local particle dynamics of the lattice-gas method.
\S\ref{long-range-2-body-interactions} describes the long-range lattice-gas and
offers a simple
theoretical result in the Boltzmann limit.  Finally,
\S\ref{simulation-results} presents some
numerical results obtained with the model.  A closing discussion of the main
points of this paper
is given in \S\ref{discussion}.  The appendix contains a formal construction of
a long-range
lattice-gas with a single species of particles obeying detailed balance and
shows its limitations.

\section{Lattice-Gas Automata}
\label{lattice-gas-automata}

An extremely abridged description of local lattice-gas dynamics is given here
since detailed
descriptions can be found elsewhere \cite{wolfram-86,frisch-87}.
Particles, with mass $m$,  propagate on a spacetime lattice with $N$ spatial
sites, unit cell size
$l$, time unit $\tau$, with speed $c=\frac{l}{\tau}$. A particle's state is
completely specified at
some time, $t$,  by specifying its position on the lattice, $\x$, and its
momentum, $\p=mc \ea$
with lattice vectors $\ea$ for $a=1,2,\dots,B$. The particles obey Pauli
exclusion since only one
particle can occupy a single state at a time.  The total number of
configurations per site is
$2^{B}$. The total number of states available in the system is $2^{B N}$.  The
lattice-gas cellular
automaton equation of motion is

\begin{equation}
n_a (\x+l\ea, t+\tau) = n_a(\x,t) + \Omega_a ( \vec{n}(\x,t) ) ,
\end{equation}
where the particle occupations and collisions are denoted by $n_a$ and
$\Omega_a$,
respectively.

For a two-dimensional hexagonal lattice, the spatial coordinates of the lattice
sites may be
expressed as follows

$\x_{ij} = \left( \frac{\sqrt{3}}{2} j , i - \frac{1}{2} \bmod_2j \right)$
where $i$ and $j$ are rectilinear indices that specify the memory array
locations used to store
the lattice-gas site data. Given a particle at site $(i,j)$, it may be shifted
along vector $\vec{\bf r}
= r\ea$ to a remote site $(i',j')_a$ by the following mapping:

$\left( i+\frac{r+1}{2}-\bmod_2j \bmod_2r ,  j \mp r \right)_{1,4}$,

$\left( i-\frac{r}{2}-\bmod_2j\bmod_2r ,  j \mp r \right)_{2,5}$,

$\left( i\mp r ,  j  \right)_{3,6}$. These streaming relations are equivalent
to memory address
offsets.  The modulus operator is base 2 because even and odd rows must be
shifted as a
hexagonal lattice is embedded into a square lattice.

\section{Long-Range 2-Body Interactions}
\label{long-range-2-body-interactions}

\begin{figure}[htb]
\hspace{1.0in}\epsfxsize=4.0in
\epsffile{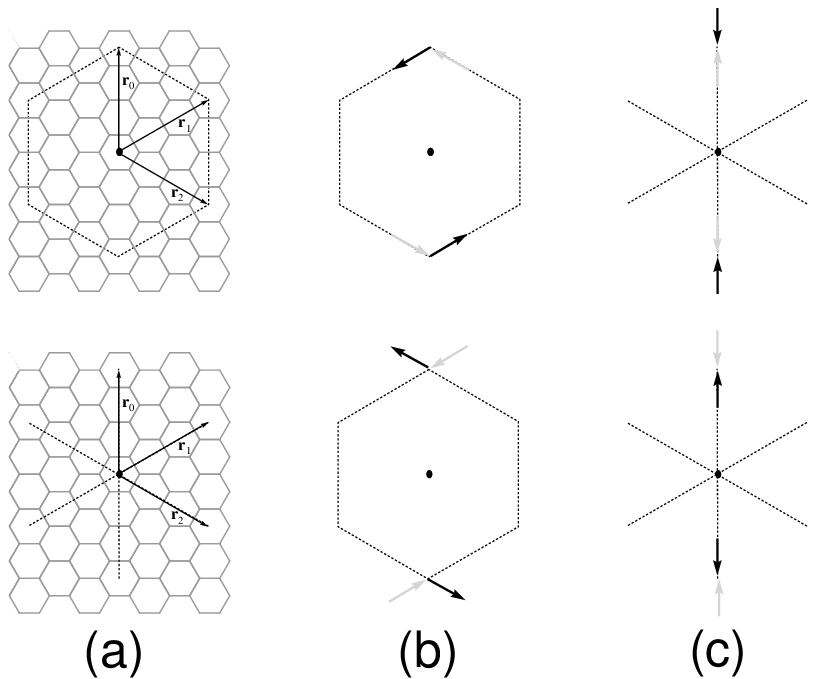}
\caption[Long-Range Bound States]{(a) Bound state orbits where the dotted path
indicates the
particle's closed trajectory; (b) $|\Delta p| = 1$ with one unit of angular
momentum, a
counter-clockwise attractive collision and its repulsive conjugate; and (c)
$|\Delta p| = 2$ with
zero angular momentum collision conjugates.  Not included in the figure are the
time-reversed
partners of (b).}
\label{long-range-orbits}
\end{figure}
An interparticle potential, $V(\x-\x')$, acts on particles spatially separated
by a fixed distance,
$\x-\x' = 2\r$.   An effective interparticle force is caused by a non-local
exchange of momentum.
Momentum conservation is violated locally, yet it is exactly conserved in the
global dynamics.

	For the case of an attractive interaction, there exists a bound states in
which two particles
orbit one another.   Since the particle dynamics are constrained by a
crystallographic lattice we
expect polygonal orbits.  In figure~\ref{long-range-orbits}a
we have depicted two such orbits for a hexagonal lattice-gas. The radius of the
orbit is $r$.
Two-body finite impact parameter collisions are depicted in
figures~\ref{long-range-orbits}b and
\ref{long-range-orbits}c. Momentum exchanges occur along the principle
directions.  The
time-reversed partners of the collisions in figures~\ref{long-range-orbits}b
are included in the
model. The interaction potential is not spherically symmetric, but has an
angular anisotropy.  In
general, it acts only on a finite number of points on a shell of radius $r$.
The number of lattice
partitions necessary per site is half the lattice coordination number, since
two particles lie on a
line. Though microscopically the potential is anisotropic, in the continuum
limit obtained after
coarse-grain averaging, numerical simulation done by Appert, Rothman, and
Zaleski indicates
isotropy is recovered \cite{appert-91}.

Constraint equations\footnote{We are simplifying this development by assuming a
single speed
lattice-gas. Consequently we do not have to explicitly write a term to conserve
energy since
here energy conservation follows by default.} for momentum  conservation and
parallel and
perpendicular momentum exchange are respectively
\begin{eqnarray}
\label{momentum-conservation}
\e_\alpha - \e_\beta + \e_\mu - \e_\nu & = & 0 \\
\label{momentum-exchange-parallel}
\left(\e_\alpha - \e_\beta - \e_\mu + \e_\nu\right)\cdot \r  & = & 2\Delta p \\
\label{momentum-exchange-perpendicular}
\left(\e_\alpha - \e_\beta - \e_\mu + \e_\nu\right)\times \r  & = & 0
\end{eqnarray}
where $\Delta p$ is the momentum change per site due to long-range collisions.
The sum and
difference of (\ref{momentum-conservation}) and
(\ref{momentum-exchange-parallel}) reduce to

\begin{equation}
\label{site-momentum-change}
(\e_\alpha)_y - (\e_\beta)_y  =  \Delta p \;\;\;\;\;\;
(\e_\mu)_y - (\e_\nu)_y   =  -\Delta p.
\end{equation}
The possible non-zero values of a site's momentum change may be $\Delta p =
\pm 1$ and
$\pm 2$.  As mentioned, the cases for $\Delta p < 0 $ led to bound states with
angular
momentum 0 and 1.  To satisfy (\ref{site-momentum-change}),  consider the case
where
$(\e_\alpha)_y = -(\e_\mu)_y$ and $(\e_\beta)_y = -(\e_\nu)_y$.
\footnote{Alternatively one
could have chosen $(\e_\alpha)_y = (\e_\nu)_y$ and $(\e_\beta)_y =
(\e_\mu)_y$.} The
possible collisions where $\hat{r}=\e_3$ are depicted in
figure~\ref{long-range-orbits}.

The reversible interactions are 2-body collisions with a finite impact
parameter of $2r$.

For $r=0$, they reduce to the 2-body collisions in the FHP lattice-gas:
the $|\Delta p| = 1$ collisions reduce to $\pm \frac{2\pi}{3}$ rotations of
momenta states, and the
$|\Delta p| = 2$ collisions reduce to the identity operation.

\begin{figure}[htb]
\hspace{1.0in}\epsfxsize=4.0in
\epsffile{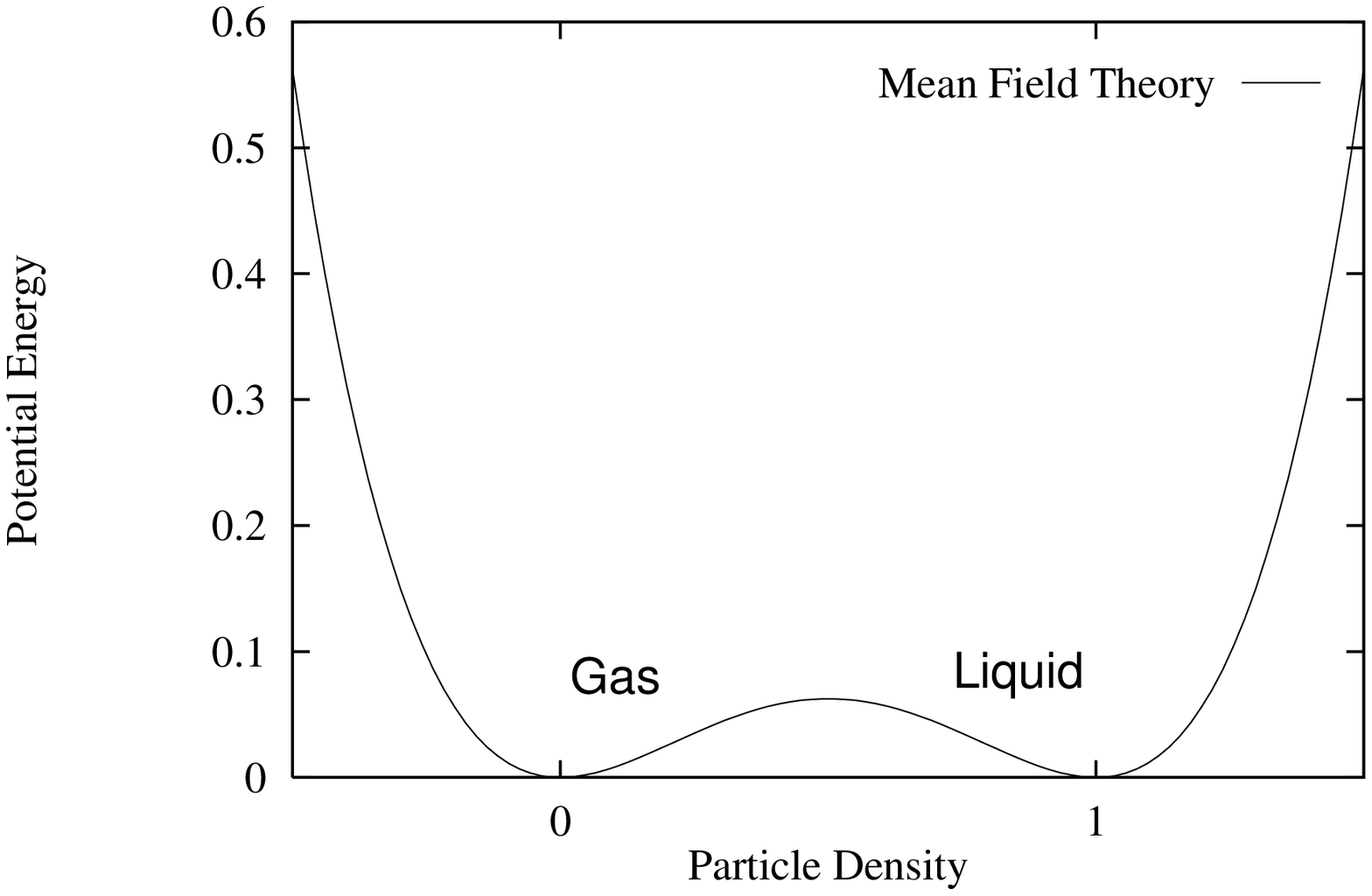}
\caption[Potential Energy versus Density at Zero Temperature]{Potential energy
versus particle
density in the zero temperature limit.  Letting $d \rightarrow
\psi-\frac{1}{2}$, then $V(\psi) =
\frac{1}{16}- \frac{1}{2}\psi^2 + \psi^4$ which has a Landau-Ginsburg form.}
\label{lr_potential_energy}
\end{figure}
Let $V[d(x)]$ represent the potential energy due to long-range interactions,
where $d(x)$ is the
probability of finding a particle at position $x$.  For a 2-body interaction to
occur at $x$ and $x'$,
one must count the chance of having two particles and two holes at the right
locations, so in the
Boltzmann limit one may write a probability of collision, $P(x,x'; v,v')$, as
\begin{equation}
P(x,x'; v,v') = d(x) (1-d(x)) d(x') (1-d(x')) \delta(|\vec{v}'-\vec{v}|)
\end{equation}
and if the system is uniformly filled, this simplifies to
\begin{equation}
P(x,x'; v,v') = d^2 (1-d)^2 \delta_{v'v}.
\end{equation}
Letting $m$, $c$, $r=|x-x'|$, and $l$ denote the particle mass, particle
velocity, 2-body
interaction range, and lattice cell size, one may write the potential energy as
\begin{eqnarray}
V(d) & = & \alpha mc^2  \left(\frac{r}{l}\right) P(x,x'; v,v') \\
     & = & \alpha mc^2  \left(\frac{r}{l}\right) d^2 (1-d)^2 \delta_{v'v},
\end{eqnarray}
where the value of the coefficient $\alpha$ depends on the magnitudes of
momenta exchanged.
Here d ranges from 0 to 1, and is just the particle filling fraction.  $V(d)$
has two minima, at
$d=0$ and $d=1$, see figure~\ref{lr_potential_energy}.

\begin{figure}[htb]
\hspace{1.5in}\epsfxsize=3.0in
\epsffile{figures/dp=1_heat_bath_collisions.eps}
\caption[Particle Coupling to the Heat Bath]{Examples of long-range collisions
that locally
conserve mass, momentum, and energy. $|\Delta p| = \pm 1$ interactions along
the
$\r_0$-direction coupled to a heat bath: (a) attractive case; and (b) its
adjoint, repulsive case.
Transitions probabilities for these collisions have a Monte Carlo form as they
are biased by the
density of heat-bath particles, rendered here with wavy lines.  Head of the
gray arrows indicates
particles entering the sites at $\r_0$ and $-\r_0$ at time $t$.  Tail of the
black arrows indicates
particles leaving those sites at time $t+\tau$.}
\label{dp=1_heat_bath_collisions}
\end{figure}

One may consider a slightly more complicated interaction, where the 2-body
collisions are
coupled to a second kind of particle whose filling fraction is denoted by $h$.
In the slightly more
complicated interaction, the form of $V(d)$ given still holds, but only for
$h=0$.  Here is the
slightly more complicated version of things.  Letting $m=c=\frac{r}{l}=1$, the
complete form of
the interaction energy is
\begin{equation}
V(d,h) =  d^2 (1-d)^2 (1-h)^2 - d^2 (1-d)^2 h^2.
\end{equation}
The first term is two $d$'s transitioning to a lower configurational energy
state and thus emitting
two $h$'s to conserve energy.  The second term is two $d$'s transitioning to a
higher
configurational energy state by absorbing two $h$'s.  Local conservation of
momentum and
energy is recovered. For convenience we write
\begin{equation}
\label{potential-energy}
V(d,h) = d^2 (1-d)^2 (1-2h).
\end{equation}
{}From (\ref{potential-energy}), $V(d,h) = 0$ for $h = \frac{1}{2}$. Since the
hath-bath particles are
fermi-dirac distributed, the effective temperature is $k_B T =
\varepsilon_o(\log
\frac{1-h}{h})^{-1}$, and $h = \frac{1}{2}$ corresponds to $T=\infty$ and $h =
0$ corresponds to
$T=0$.  Numerical simulation corroborates this.

The pressure, $p$, in the gas is written
\begin{equation}
\label{nonideal-pressure}
p(d,h) = c_s^2 d + V(d,h)
\end{equation}
where $c_s < c$ is the sound speed.  This is the non-ideal equation of state
that is responsible
for the liquid-gas phases observed in numerical simulations of this system.

\section{Simulation Results}
\label{simulation-results}

\begin{figure}[htb]
\hspace{1.0in}\epsffile{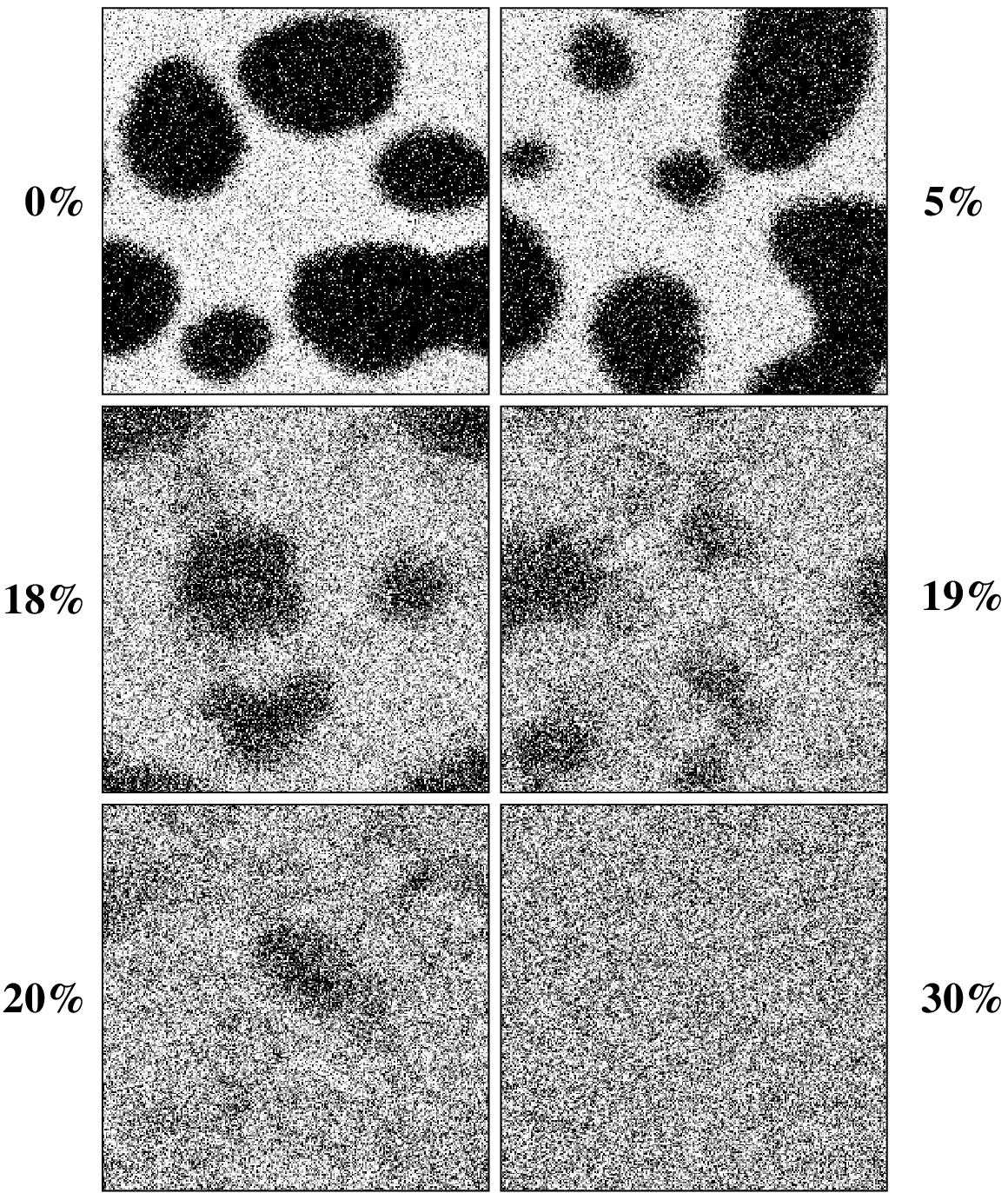}
\caption[Lattice-Gas Configurations Verses Effective Temperature]{Several
lattice-gas
configurations obtained after 500 iterations starting from random initial
$256\times 256$
configurations with 30\% particle filling. The six configuration are coupled to
a heat-bath with 0,
5, 18, 19, 20, and 30\% heat bath filling.}
\label{temp-liq-gas}
\end{figure}
\begin{figure}[htb]
\hspace{0.5in}\epsffile{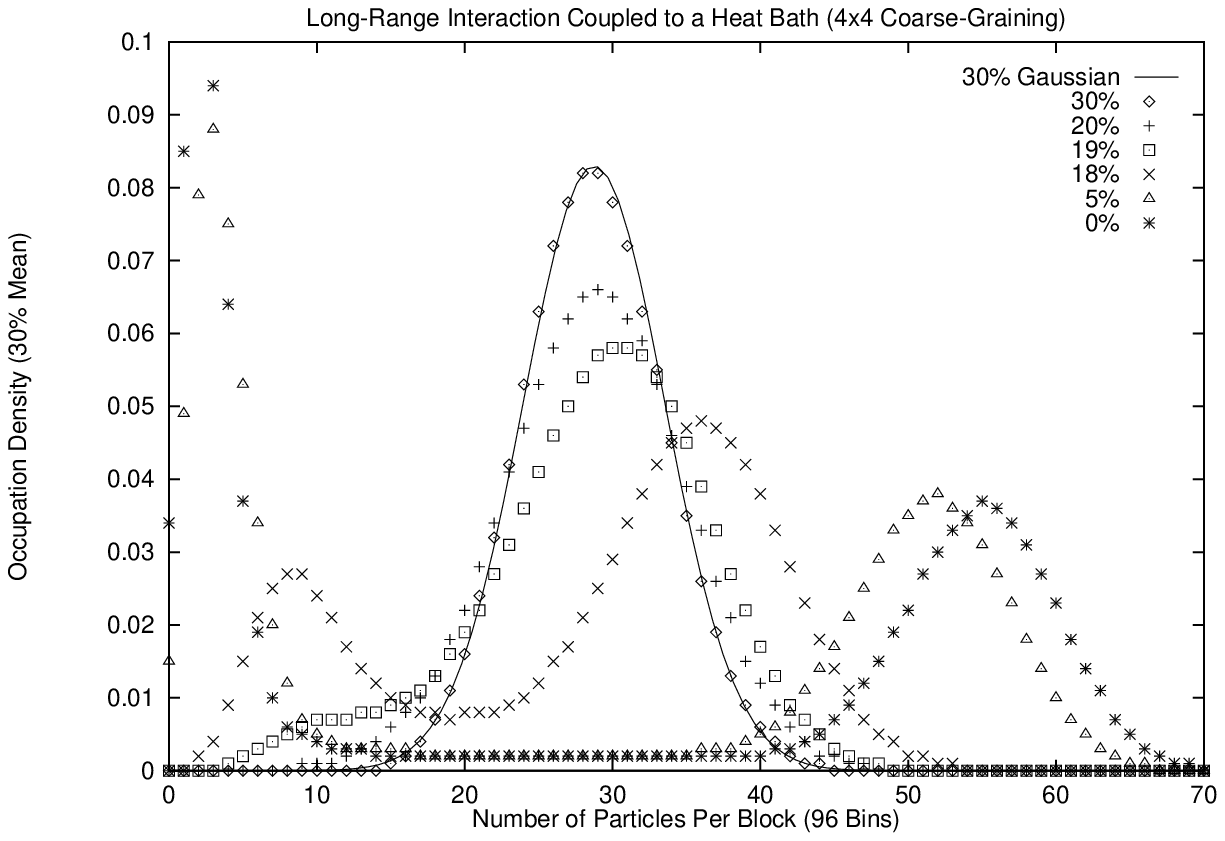}
\caption[Mass Frequency Distributions Verses Effective Temperature]{Mass
frequency
distributions obtained by course-grained averaging over the lattice-gas number
variables.
$4\times 4$ blocking is used on a $256\times 256$ hexagonal lattice. Result for
particle filling
fraction of 0.3.}
\label{multidist.30.256}
\end{figure}

The dynamics of the model in contact with a fixed temperature heat bath is
tested by numerical
simulation.  A coarse-grained mass frequency distribution is measured after the
system has
evolved for a fixed amount of time.  The lattice-gas is initialized with a
random configuration and
allowed to evolve for 500 time steps for several bath filling fractions: 30\%
to  20\%, 19\%, 18\%,
5\%, and 0\%.  Resulting system snapshots are illustrated in
figure~\ref{temp-liq-gas}.

If the lattice-gas is above the transition temperature, the particles are
uniformly spread over the
lattice.  As the system evolves while in contact with a finite temperature heat
bath
coarse-grained $4\times 4$ block averages over a $256\times 256$ lattice are
taken over the
lattice-gas number variables to produce a mass frequency distribution for a
large number of
temperatures and the liquid and gas densities are found.  A mass frequency
distribution obtained
by this coarse-grained block averaging procedure is a Gaussian with its mean
located exactly at
the initial particle density.  A normalized Gaussian fit, centered at particle
density 0.3, is shown
in figure~\ref{multidist.30.256} as is cross-section plots of the distribution
at different
temperatures.  As the temperature decreases, the distribution widens and
becomes bimodal.
The mean of the low density peak gives the gas phase density and the mean of
the high density
peak gives the liquid phase density.  The order parameter for the liquid-gas
transition is the
difference of the liquid and gas densities, $\psi = \rho_L - \rho_G$
\cite{stanley-71}.

\begin{figure}[htb]
\hspace{0.5in}\epsfxsize=5.0in
\epsffile{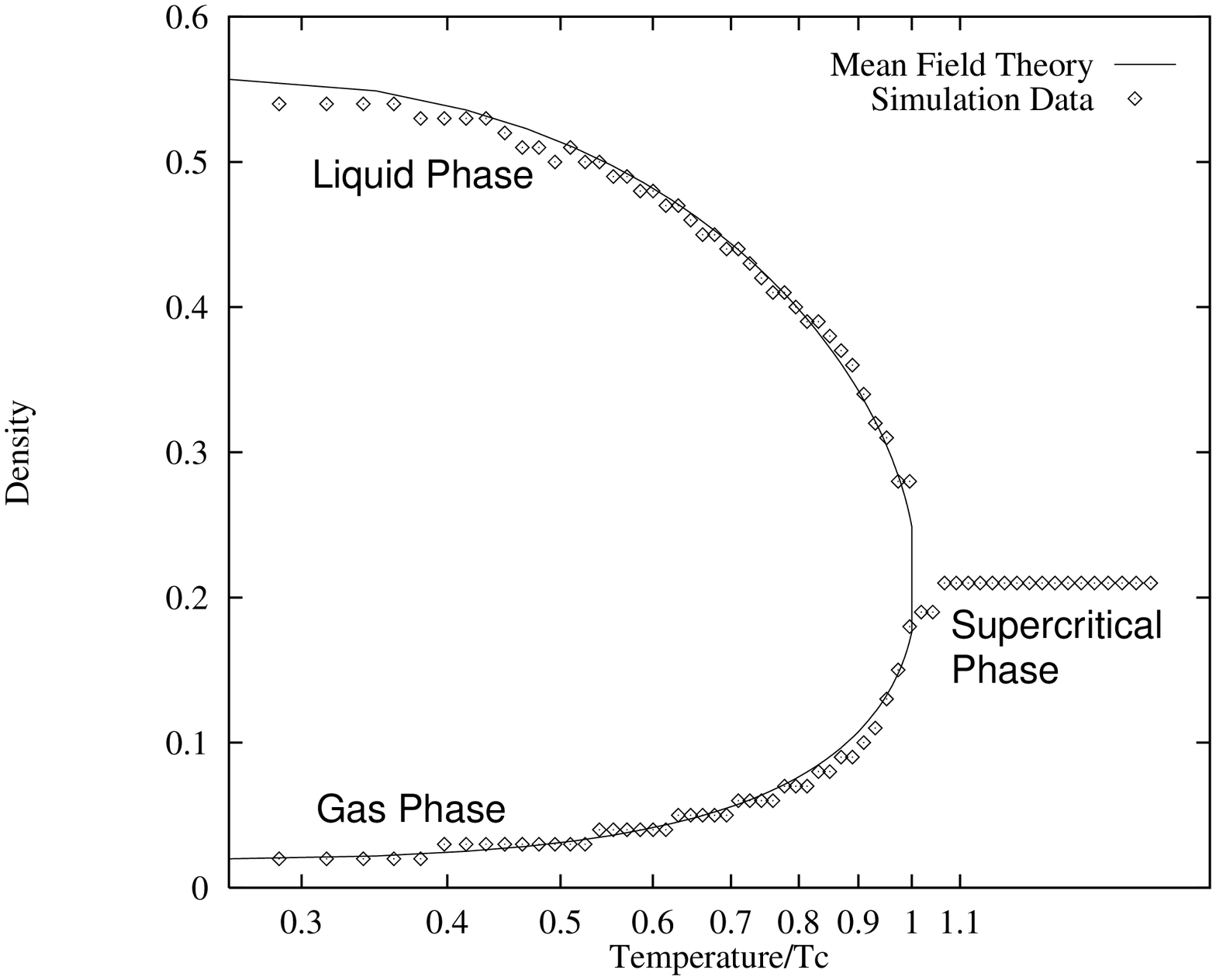}
\caption[Liquid-gas Coexistence Curve]{Liquid-gas coexistence curve determined
from the
particle mass frequency distribution for different heat bath temperatures.}
\label{lr_coexist}
\end{figure}

If the heat-bath temperature is held constant, the dynamics is no longer
reversible. The ordered
phase persists and the simulation method becomes like a  Monte Carlo Kawasaki
updating
scheme ({\it i.e.} the exchange of randomly chosen spins).  However, using this
long-range
interaction method, momentum is exactly conserved and kinetic information
retained.
Therefore, the dynamical evolution of the finite temperature multiphase system
is accessible,
even near the critical temperature.   Figure~\ref{lr_coexist} shows a
comparison of numerical
simulation data obtained by this procedure to an analytical calculation done in
the Boltzmann
limit by analytically carrying out a Maxwell construction.  The Gibbs free
energy of the
lattice-gas can be written analytically since the pressure's dependence on
density and
temperature is known, (\ref{nonideal-pressure}).  A Maxwell construction then
correctly predicts
the liquid and gas phase densities at any given heat-bath temperature.  This
mean field type of
calculation itself is very interesting and is a good example of how analytical
calculations are
possible in simple discrete physical models like lattice-gases. The result
shown in
figure~\ref{lr_coexist} is similar to the order parameter curve, magnetization
versus temperature,
of an Ising model.  Spin up, $\langle M\rangle=+1$,  and spin down, $\langle
M\rangle=-1$,
domains are analogous to the liquid and gas phases.

\section{Discussion}
\label{discussion}

The model is a simple discretization of molecular dynamics with interparticle
potentials.
Because of the model's small local memory requirement, the dynamics of large
systems can be
implemented on a parallel architecture, as has been done on the cellular
automata machine
CAM-8 \cite{margolus-93}.

The  main points of this paper are:

1. Coupling the particle dynamics to a fixed temperature heat-bath sets the
transition
probabilities and causes a net attractive interparticle potential in the
macroscopic limit.  The
heat-bath is comprised of a set of lattice-gas particles possessing a unit of
heat. The heat bath
density, $h$, controls the heat-bath's temperature by the fermi-dirac
distribution, $k_B T =
\varepsilon_o(\log \frac{1-h}{h})^{-1}$. The system possesses a nonideal $P\rho
T$ equation of
state.  With the models in contact with a controlled heat-bath, one should
classify it as finite
temperature model with heat-bath dynamics similar to a Monte Carlo Kawasaki
updating, yet
retaining essential kinetic features.  It is a step toward a more complete
long-range lattice-gas
that preserves an interaction energy.

2. The equation of state is known in the Boltzmann limit. A Maxwell
construction predicts the
liquid-gas coexistence curve. Van der Waal coefficients can be determined to
map the
simulation on to a particular physical liquid-gas system.

It is hoped that a lattice-gas model, of the kind presented here, will become a
valuable new tool
for analytically and numerically studying the dynamic critical behavior of
multiphase systems.

\section{Acknowledgements}

I would like to acknowledge Dr Norman Margolus and Mark Smith of the MIT
Laboratory for
Computer Sciences, Information Mechanics Group for their suggestion of
considering reversible
dynamics, a topic they have considered for some time in their investigation of
cellular automata
models of physics.  And I would like to especially thank them for the very many
interesting
discussions on this topic, and in particular those concerning multiphase
systems.  I must also
express my thanks to Dr Stanley Heckman for his review of this paper.

\appendix

\section{Reversibility-Neutrality Statement}

The main thrust of this paper has been to describe a momentum conserving
multiphase
lattice-gas.  However, the lattice-gas dynamics is not reversible.  In this
appendix a reversible,
momentum conserving lattice-gas with long-range interactions is presented.
However, when
this model is coupled to a heat-bath, there is an imperfect tracking of
interaction energy. This is
a limitation of a reversible lattice-gas with a single particle species.  An
improved version of this
reversible lattice-gas model with long-range interactions could be implemented
(likely with a
species of ``bound state'' particles included) so that simulations are carried
out in a
microcanonical ensemble analogous to Ising models with auxiliary demons
introduced by Creutz
\cite{creutz-prl92}, and Toffoli and Margolus \cite{toffoli-87}, but this is
not presented here.  A
formalism is presented for describing the unitary evolution of a lattice-gas
with long-range
interactions.

An unbiased reversible lattice-gas model cannot have a net attractive
interparticle potential, the
lattice-gas particles must be neutral.  Let us see why.
For any reversible computational model a unitary operator maps the
computational state at
some time to the state at the next time iteration.  This unitary matrix can be
expressed as the
exponential of a hermitian operator, a kind of computational Hamiltonian.  This
mathematical
construction is similar to a quantum mechanical description \cite{benioff-82}.

Using the notation of multiparticle quantum mechanical systems in the second
quantized
number representation \cite{fetter-71}, all states of the system are enumerated
sequentially by
$\alpha_x = 1 \dots N_{\hbox{\tiny total}}$ where $N_{\hbox{\tiny total}} = B
N$. For each
$\alpha$ there is an associated site $\x$ that is indicated by a subscript.
Denote the vacuum
state of the system by $| 0 \rangle$ where all $n_{\alpha_x} = 0$ for all
states $\alpha$.  Using
creation and annihilation operators $\ahd_{\alpha_x}$ and $\ah_{\alpha_x}$ to
respectively
create and destroy a particle in state $\alpha$ at lattice position $\x$, any
arbitrary system
configuration $\psi$ with $P$ particles can be formed by their successive
application on the
vacuum
\begin{equation}
| \psi  \rangle = \prod_{p=1}^P \ahd_{\alpha_p}  | 0 \rangle ,

\end{equation}
where particle one is  in state $\alpha_1$ at lattice node $\x_1$, particle two
is  in state
$\alpha_2$ at lattice node $\x_2$, etc.

The number operator is $\hat{n}_{\alpha_x} = \ahd_{\alpha_x} \ah_{\alpha_x}$.
To completely specify the dynamics the anticommutation relations are required.
 Since there is
exclusion of boolean particles at a single momentum state, we have
\begin{eqnarray}
\label{anticommutators-boolean-local}
\{\ah_{\alpha}, \ahd_{\alpha} \}  & =  & \one \\
\{\ah_{\alpha},  \ah_{\alpha} \}  & =  & 0 \\
\{\ahd_{\alpha}, \ahd_{\alpha}\}  & =  & 0.
\end{eqnarray}
However, the boolean particles are completely independent at different momentum
states, and
so the operators commute
\begin{eqnarray}
\label{anticommutators-boolean-remote}
\left[\ah_{\alpha}, \ahd_{\beta}\right]   & =  & 0 \\
\left[\ah_{\alpha},  \ah_{\beta}\right]   & =  & 0 \\
\left[\ahd_{\alpha}, \ahd_{\beta}\right]  & =  & 0
\end{eqnarray}
for $\alpha\neq\beta$.

A unitary evolution operator that describes the complete evolution of the
lattice-gas may be
partitioned into a streaming and collisional part,  $\U^{\tiny\hbox{o}}$ and
$\U^{\tiny\hbox{int}}$
respectively.  The full system evolution operator is a product of these two
operators
$\U = \U^{\tiny\hbox{o}}\U^{\tiny\hbox{int}}$.
The operator $\U^{\tiny\hbox{o}}$ is constructed using a unitary single
exchange operator
denoted by $\mex^{(1)}_{\alpha\beta'}$.  All permutations of single boolean
particle states may
be implemented by successive application of this momentum-exchanger.
We will use the same symbol, $\mex^{(1)}_{\alpha\beta'}$, to denote the
permutations between
state $\alpha$ at site $\x$ and states $\beta'$ at site $\x'$.  We wish to
construct $\mex^{(1)}$
from the boolean lattice-gas creation and annihilation operators.

We require that $\mex^{(1)}$ is unitary, $(\mex^{(1)})^2=\one$, that it
conserve the number of
particles, $[\mex^{(1)},{\hat N}]=0$, and that $\mex^{(1)} \vac = \vac$. It has
the form
\begin{equation}
\label{momentum-exchanger-single-boolean}
\mex^{(1)}_{\alpha \beta'} = \ahd_\alpha \ah_{\beta'} +

\ahd_{\beta'} \ah_{\alpha} + \one -
\ahd_\alpha \ah_\alpha \ah_{\beta'} \ahd_{\beta'}-
\ahd_{\beta'} \ah_{\beta'}\ah_\alpha \ahd_\alpha.
\end{equation}
This can be written in the form
\begin{equation}
\label{mex1-simple-boolean}
\mex^{(1)}_{\alpha \beta'} = \one  - 2 \N^{(1)}_{x x'} = e^{i \pi {\hat
N}^{(1)}_{x x'}}.
\end{equation}
For a set of $N$ states, $\{1,2,\dots,N-1,N\}$, two $C_{N}$ rotation operators
can be
implemented
\begin{eqnarray}
\hat R^{C_N} & = & \mex^{(1)}_{N-1,N} \mex^{(1)}_{N-2,N-1} \cdots
\mex^{(1)}_{12}

= \prod_{i=1}^{N-1} \mex^{(1)}_{N-i, N-i+1}
\\
\hat R^{C_N^{N-1}} & = & \mex^{(1)}_{12} \mex^{(1)}_{23} \cdots
\mex^{(1)}_{N-1,N}
= \prod_{i=1}^{N-1} \mex^{(1)}_{i,i+1}
\end{eqnarray}
that are rotations by $\pm \frac{N}{360}^\circ$.  Suppose we pick a subspace to
be the set of
states, ${{\cal P}_a}$, with momentum $mc\e_a$.  Then following our
construction, we have
found a method to implement a unitary streaming operator, $\stream_a$, along
direction-$a$
\begin{equation}
\stream_{a} = \prod_{\{\alpha,\beta\}\in{\cal P}_a} \mex^{(1)}_{\alpha_{x}
\beta_{x+le_a}}.
\end{equation}
The free part of the evolution is then simply
\begin{equation}
\U^{\tiny\hbox{o}} = \prod_{a=1}^{B} \stream_a = \prod_{a=1}^{B}
\prod_{\{\alpha,\beta\}\in{\cal
P}_a} \mex^{(1)}_{\alpha_{x} \beta_{x+le_a}}.
\end{equation}
The corresponding kinetic energy part of the Hamiltonian to leading order is

\begin{equation}
\label{qc-hamiltonian-free}
\ham^{\tiny\hbox{o}}  =  \sum_{\langle x x' \rangle}

\ahd_{\alpha_x} \ah_{\beta_{x'}} + \ahd_{\beta_{x'}} \ah_{\alpha_{x}}  + \cdots
\end{equation}
where the sum is over all bonds of the lattice taken over partitions along
principle lattice
directions and where for brevitity the following short-hand notation is used:
$\prod_{a=1}^{B}
\prod_{\{\alpha,\beta\}\in{\cal P}_a} \rightarrow \sum_{\langle x x' \rangle}$
when  $x + l e_a
\rightarrow x'$.

The operator $\U^{\tiny\hbox{int}}$ is constructed using a unitary double
exchange operator,
denoted by $\mex^{(2)}_{\alpha\beta\mu'\nu'}$, a generalization of the single
exchange
operator. All permutations of two boolean particles may be implemented by
successive
application of this momentum-exchanger,  where the permutations for particle
one occurs
between states $\alpha$ and $\beta$ at site $\x$ and for particle two between
states $\mu'$ and
$\nu'$ at site $\x'$.

We require that $\mex^{(2)}$ is unitary, $(\mex^{(2)})^2=\one$, that it
conserve the number of
particles, $[\mex^{(2)},{\hat N}]=0$, and that $\mex^{(2)} \vac = \vac$.
A relation identical to (\ref{mex1-simple-boolean}) exists for the double
boolean exchange
operator
\begin{equation}
\label{mex-simple-double-boolean}
\mex^{(2)}_{\alpha\beta\mu'\nu'} = \one  - 2 \N^{(2)}_{x x';v v'}  = e^{i \pi
{\hat N}^{(2)}_{x x'; v
v'}}.
\end{equation}

Let us assume we have a two-particle state $|\gamma \sigma'\rangle =
\ahd_{\gamma_x}\ahd_{\sigma_{x'}} \vac$, where $\gamma = \alpha$ or $\beta$,
and $\sigma' =
\mu'$ or $\nu'$.  It has the form
\begin{equation}
\label{momentum-exchanger-double-boolean}
\mex^{(2)}_{\alpha\beta\mu'\nu'} =

\ahd_\alpha \ah_\beta \ahd_{\mu'} \ah_{\nu'} +

\ahd_\beta \ah_\alpha \ahd_{\nu'} \ah_{\mu'} +

\one -
\ahd_\beta \ah_\beta \ahd_{\nu'} \ah_{\nu'} \ah_\alpha \ahd_\alpha \ah_{\mu'}
\ahd_{\mu'}-

\ahd_\alpha \ah_\alpha \ahd_{\mu'} \ah_{\mu'}  \ah_\beta \ahd_\beta \ah_{\nu'}
\ahd_{\nu'}.
\end{equation}
Suppose we pick a subspace to be the set of states, ${{\cal V}_a}$, where
moment exchanges
$\pm mc\e_a$ can occur between two particle pairs.  Then a unitary collision
operator, $C_a$,
in this subspace (with momentum exchanges along a principle lattice direction)
is
\begin{equation}
C_a = \prod_{\{\alpha,\beta\,\mu,\nu\}\in{\cal V}_a} \mex^{(2)}_{\alpha_{x+r
e_a} \beta_{x+r
e_a}\mu_{x-r e_a} \nu_{x-r e_a}}.
\end{equation}
The interaction part of the evolution is then simply
\begin{equation}
\U^{\tiny\hbox{int}} = \prod_{a=1}^{\frac{B}{2}} C_a =
\prod_{a=1}^{\frac{B}{2}}
\prod_{\{\alpha,\beta\,\mu,\nu\}\in{\cal V}_a} \mex^{(2)}_{\alpha_{x+r e_a}
\beta_{x+r
e_a}\mu_{x-r e_a} \nu_{x-r e_a}}.
\end{equation}
The corresponding potential energy part of the Hamiltonian to leading order is

\begin{equation}
\label{qc-hamiltonian-int}
\ham^{\tiny\hbox{int}}  =  \sum_{\langle x x' ; v v'\rangle}

\ahd_{\alpha_x} \ah_{\beta_x} \ahd_{\mu_{x'}} \ah_{\nu_{x'}} +

\ahd_{\beta_x} \ah_{\alpha_x} \ahd_{\nu_{x'}} \ah_{\mu_{x'}} + \cdots ,
\end{equation}
where for brevity the following short-hand notation is used:
$\sum_{a=1}^{\frac{B}{2}}
\sum_{\{\alpha,\beta\,\mu,\nu\}\in{\cal V}_a} \rightarrow \sum_{\langle x x' ;
v v' \rangle}$ when
$x + r e_a \rightarrow x$ and $x - r e_a \rightarrow x'$.
(\ref{qc-hamiltonian-free}) and
(\ref{qc-hamiltonian-int}), imply to leading order, the full lattice-gas
Hamiltonian will have the
form
\begin{equation}
\label{hamiltonian}
\ham = \ham^{\tiny\hbox{o}} + \ham^{\tiny\hbox{int}} + \cdots = \sum_{\langle x
x'\rangle}
\ahd_{\alpha_x} \ah_{\beta_{x'}} + \sum_{\langle x x' ; v v'\rangle}
\ahd_{\alpha_{x}} \ah_{\beta_{x}} \ahd_{\mu_{x'}} \ah_{\nu_{x'}} +

c.c. + \cdots
\end{equation}
The complex conjugate terms arise in (\ref{hamiltonian}) because of
reversibility and ensure the
hermiticity of the Hamiltonian.  If the interaction term in the Hamiltonian
covers all possible
attractive interactions, its complex conjugate will then cover all possible
repulsive interactions.
Since (\ref{hamiltonian}) is a completely general way of specifying any set of
2-body collisions,
and since it necessarily describes invertible lattice-gas dynamics because of
the unitarity of the
evolution operator, we have a general proof that an unbiased reversible
lattice-gas model
cannot have a net attractive interparticle potential and that the lattice-gas
particles must be
neutral.

\begin{figure}[htb]
\hspace{0.5in}\epsffile{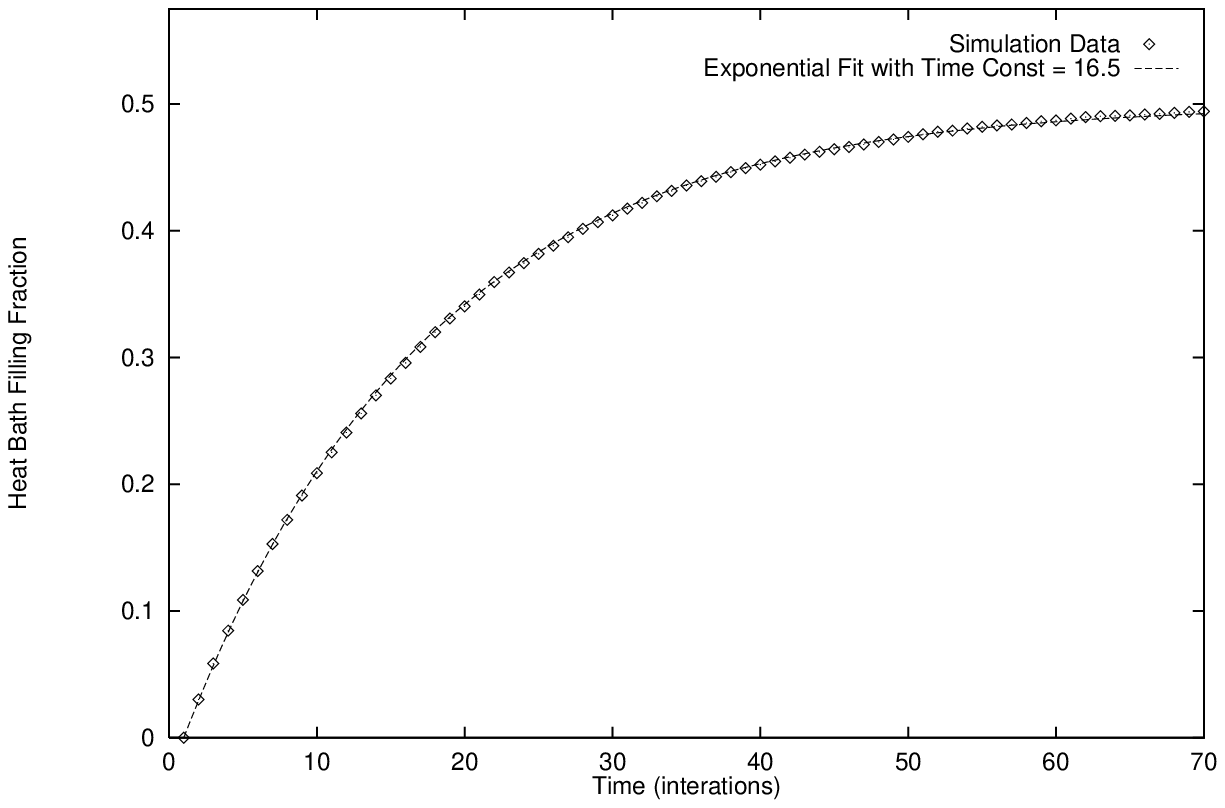}
\caption[Transient Behavior of a Reversible Lattice-Gas]{Transient behavior of
a reversible
lattice-gas with long-range interactions.  Particle density at approximately
$\frac{1}{6}$-filling. An
exponential increase to half-filling with a time constant of 16.5$\tau$ is
observed.}
\label{pdensity.16.128}
\end{figure}

As the dynamics is reversible, the system quickly moves to a maximal entropy
state where the
net attractive interparticle potential vanishes.

The liquid-gas coexistence phase may persist indefinitely given a net
attractive interaction.  In a
reversible system a net attraction exists for a short while, only so long as
most heat bath states
are not populated.  Once the heat bath gains a significant population, only the
local interactions
remain and consequent diffusion drives the system back to a disordered phase.
The maximal
entropy state of the heat bath occurs at half-filling, $h = \frac{1}{2}$,  and
consequently at this
heat bath density it cannot encode any more information about heating from the
lattice-gas so
the effect of the long-range interaction must become non-existent.  Note that
is consistent with
(\ref{potential-energy}), since $V(d,h)=0$ for $h = \frac{1}{2}$. The
uncontrolled heat bath
population exponentially approaches its maximal entropy state, see
figure~\ref{pdensity.16.128},
starting from a density initially zero;  $h (t) = \frac{1}{2}(1-e^{-t/\tau})$
with the observed time
constant, $\tau=16.5$, obtained by fitting.  The time constant, $\tau$, can be
increased by
raising the number of heat bath states.

\end{document}